\documentclass[journal=jacsat,manuscript=article]{achemso}

\usepackage{chemformula} 
\usepackage[T1]{fontenc} 
\usepackage{amssymb}
\usepackage{bm}
\usepackage{amsmath}
\usepackage{accents}
\usepackage{booktabs}
\usepackage{dcolumn}
\usepackage{chemfig}

\author{Sobin Alosious}
\affiliation{Department of Applied Mechanics, Indian Institute of Technology
Madras, Chennai 600036, India}
\alsoaffiliation{Department of Mathematics, School of Science, Computing and Engineering Technologies, Swinburne University
of Technology, Melbourne, Victoria 3122, Australia}
\author{Sridhar Kumar Kannam}
\affiliation[Aus]{Department of Mathematics, School of Science, Computing and Engineering Technologies, Swinburne University
of Technology, Melbourne, Victoria 3122, Australia}

\author{Sarith P. Sathian}
\affiliation[India]{Department of Applied Mechanics, Indian Institute of Technology
Madras, Chennai 600036, India}
\author{B.D. Todd}
\affiliation[Aus]{Department of Mathematics, School of Science, Computing and Engineering Technologies, Swinburne University
of Technology, Melbourne, Victoria 3122, Australia}
\email{btodd@swin.edu.au}
\title[An \textsf{achemso} demo]
  {Effects of electrostatic interactions on Kapitza resistance in hexagonal boron nitride-water interfaces}

\abbreviations{IR,NMR,UV}
\keywords{American Chemical Society, \LaTeX}

\begin{document}

%
%
%
%
%

\begin{abstract}
Electrostatic interactions in nanoscale systems can influence the heat transfer mechanism and interfacial properties. This study uses molecular dynamics simulations to investigate the impact of various electrostatic interactions on the Kapitza resistance ($R_k$) on a hexagonal boron nitride-water system.
The Kapitza resistance at hexagonal boron nitride nanotube (hBNNT)-water interface reduces with an increase in diameter of the nanotube due to more aggregation of water molecules per unit surface area. An increase in the partial charges on boron and nitride caused the reduction in $R_k$. With the increase in partial charge, a better hydrogen bonding between hBNNT and water was observed, whereas the structure and order of the water molecules remain the same. Nevertheless, the addition of NaCl salt into water does not have any influence on interfacial thermal transport. $R_k$ remains unchanged with electrolyte concentration since the cumulative Coulombic interaction between the ions, and the hBNNT is significantly less when compared with water molecules. Furthermore, the effect of electric field strength on interfacial heat transfer is also investigated by providing uniform positive and negative surface charges on the outermost hBN layers.  $R_k$ is nearly independent of the practical range of applied electric fields and decreases with an increasing electric field for extreme field strengths until the electro-freezing phenomenon occurs. The ordering of water molecules towards the charged surface leads to an increase in the layering effect, causing the reduction in $R_k$ in the presence of an electric field.

\end{abstract}
\maketitle
\section{Introduction}
A profound understanding of the fundamental mechanisms underlying nanoscale interfacial thermal transport is essential to improve the performance of various nanoscale devices across a wide range of technologies. \cite{pop2010energy,karniadakis2006microflows,cahill2003nanoscale}. 
Thermal transport and resistance at interfaces are much more significant in nanoscale systems than bulk systems since the interface can dramatically affect the observed properties of various nanoscale materials \cite{li2019anisotropic}. Carbon nanotubes (CNTs) and boron nitride nanotubes (BNNTs) have received much interest as novel two-dimensional materials in the last decade because their fascinating physical properties make them promising candidates for various nanofluidic applications. \cite{lyderic2010nanofluidics,wang2021abnormal,shannon2010science,suk2008fast,wang2010recent}
Furthermore, boron nitride materials have outperformed graphene and carbon nanotubes in some features such as electrical insulation, chemical inertness, and biological compatibility.
\cite{blase1994stability,zhi2010boron,chen2009boron}
Molecular dynamics (MD) simulation techniques are widely used to compute various properties of boron nitride systems that require a well-defined force field and water-BN interaction parameters. Multiple studies have been conducted to develop the BN force field parameters with the help of different techniques such as diffusion Monte Carlo (DMC), random phase approximation (RPA) and density functional theory (DFT) \cite{wu2016hexagonal,won2008structure}.
With the help of these force field parameters, researchers have extensively studied different mechanisms in hexagonal boron nitride systems such as ion selectivity, wetting behaviour and water transport
\cite{won2007water,azamat2015removal,won2008water}. Nevertheless, compared to graphene and carbon nanotube (CNT) systems, interfacial heat transfer and thermal resistance in hexagonal boron nitride-water systems have received little attention. Uhlig \textit{et al.} \cite{uhlig2021atomically}  investigated the interfacial water structure of hydrophobic and hydrophilic surfaces using 3D-AFM imaging with atomic-scale spatial precision. They showed that the interfacial water structure on crystalline surfaces has two contrasting configurations at the atomic scale. Their findings demonstrate the complexities of real water interfaces, where a small quantity of hydrocarbons may dominate the interface's fundamental characteristics.

The properties of confined water significantly change in the presence of charged surfaces or ionic substrates due to the electrostatic interactions. The effect of surface charge decoration on the interfacial heat transport in a graphene water system was studied by Ma \textit{et al.} \cite{ma2018ordered} with the help of molecular dynamics simulations. The enhanced Coloumbic force of attraction between water and diagonally charged graphene layers causes an increase in interfacial interaction strength, leading to meagre Kapitza resistance. Furthermore, the diagonal charge decoration caused the formation of ordered water molecules close to the interface. The spectral analysis reveals that the diagonal charge decoration enhances spectral thermal conductance for all frequencies. The diagonal charge decorating on the interfacial graphene sheet improves spectral thermal conductivity at all frequencies compared to the system without charge decoration. Furthermore, the characteristic peak in the spectral thermal conductance shifts to a higher frequency. These results are consistent with prior research on the solid-liquid interface based on the LJ model. \cite{saaskilahti2016spectral}.  Wang \textit{et al.} \cite{wang2018molecular} introduced a promising technique to reduce the interfacial thermal resistance between copper (Cu) and water by providing surface charges to Cu. The surfaces charges induced an orientation alignment of water molecules and reduced the distance between the surface and the water nanolayer. The enhanced heat transfer is also due to the improved coupling of vibrational density of states between water and Cu when surface charges are employed. Wei and Luo \cite{wei2018effects} studied the transport of water and the friction coefficient in carbon nanotubes and hexagonal boron nitride nanotubes using molecular dynamics simulations. The friction coefficient in zigzag hBNNT is significantly higher than zigzag CNT due to the partial charges, which leads to local potential energy traps and additional electrostatic interaction between hBNNT and water. Moreover, for an armchair hBNNT, its atomic arrangement does not create local potential energy traps; thus, the friction coefficient is less than that for a zigzag hBNNT. Qian \textit{et al.} \cite{qian2019ultralow} reported that the interfacial thermal transport in ionic liquid (IL)-solid interfaces is dominated by the atomic structure of EDL rather than the strong solid-liquid interaction using molecular dynamics simulations. The decrease in ITR with an increase in the surface charge occurs sharply in the initial stage and slowly in the later stage. They have calculated the structure factor and geometry state for the quantitative analysis of the EDL structure near the interface. A very low ITR can be achieved by the highly ordered interfacial structure of IL, which is justified by the vibrational spectrum and frequency-dependent heat flow.

Various electrolyte solutions under nano-confinement have been widely studied due to their significance in biological and nanofluidic applications. Dewan \textit{et al.} \cite{dewan2014structure} characterized the fundamental properties of interfacial water by conducting molecular dynamics simulations of alkali chloride solutions with two kinds of idealized charged surfaces. They observed a compact layer of solvents adjacent to the surface beside the diffuse region where water orientation exhibits no layering. The type of the charge distribution is the primary deciding factor for the diffuse solvent layer depth. For a realistic model of negatively charged amorphous silica, the distribution of ions and water orientation depends on the type of cations and are not well described by a simple uniform charge distribution model. Hilder \textit{et al.} \cite{hilder2009salt} showed that a silicon nitride membrane embedded with a hexagonal boron nitride nanotube could obtain a 100\% (in principle) salt rejection at higher concentrations. A water flow rate of 10.7 water molecules per nanosecond is possible even for a concentration as high as 1 M, which has a high energy barrier. Moreover, the nanotube radius determines whether the nanotube is anion-selective or cation-selective since the water structure forms a single-file chain for smaller radius nanotubes. Prakash et al. \cite{prakash2020non} studied the flow of a symmetric electrolyte via a charged nanochannel that was exposed to an axial temperature gradient. They investigated the relative contributions of the Soret effect, thermoelectric effect, and double-layer potential in the electrical double layer for different surface charges and temperature gradients. Their findings show that temperature variations may be used to produce streaming current in a charged graphene-based nanochannel, depending on the relative impact of the Soret effect and the double layer potential.

Any structural change of water is sensitive to the electric field due to the dipole moments in water molecules. Khusnutdinoff and Mokshin \cite{khusnutdinoff2019electrocrystallization} studied the effect of electric field-induced crystalization of supercooled water confined in a graphene nanochannel. By applying an electric field of 0.5 V/\AA \ to confined water at a temperature of 268 K and density 0.94 g/cm$^3$, they observed that the water crystallizes into cubic ice with some defects. When the electric field is applied, the water dipoles become aligned, and the ordinary relaxation of the metastable state transforms to an ordered phase leading to the crystalization of water. Su and Guo \cite{su2011control} investigated the impact of the flow of single-file water molecules through CNT in the presence of an external electric field. The applied electric field changes the water molecules' orientation, disturbs the wavelike density profile, and strongly influences the water flux. They observed a critical electric field $E_c$ beyond which the electric field strength does not affect the water flux since the frequency of water dipole flipping vanishes for higher electric field strength. The electric field-induced water flow can be helpful in various biological applications and the efficient design of nano pumps. Yenigun and Barisik \cite{yenigun2019electric} studied the effect of varying electric fields on interfacial thermal transport in water confined between two parallel silicon slabs. They observed a fivefold reduction in the Kapitza length by applying an electric field due to the change in silicon and water interface energy. Moreover, the thermal conductivity of the water was slightly reduced due to the alignment of dipoles in the direction of the electric field. Further increase in the electric field about 0.53 V/nm leads to electro-freezing since the aligned dipole developed a crystalline structure. Kapitza length and thermal conductivity remain unchanged with a further increase in the electric field once the electro-freezing is formed.  

In this paper, we investigate the influence of electrostatic interactions on Kapitza resistance in a hexagonal boron nitride-water system using equilibrium molecular dynamics (EMD) simulations. The effect of hBNNT partial charge, electrolyte concentration and electric field were analyzed in cylindrical and planar hBN-water systems.  The Kapitza resistance/length calculations were carried out using our previously developed EMD method \cite{alosious2021nanoconfinement} since the NEMD method is unsuitable for the cylindrical system used in this study. The Kapitza resistance is found to be decreasing with an increase in partial charge of hBNNT and applied electric field, whereas the increase in molarity of the NaCl electrolyte does not make any impact. A summary of the EMD method, methodology, detailed analysis of the obtained results, and concluding remarks are provided in the following sections of the paper.

\section{Theory}
We have recently developed an equilibrium molecular
dynamics (EMD) method to compute the Kapitza resistance
in a cylindrical nanoconfinement system (CNT-water interface).
Here we briefly summarise the concept and the final expressions of the EMD method, which applies to planar and cylindrical systems. The detailed derivation of the theory is available in the original paper \cite{alosious2021nanoconfinement}.

The Kapitza resistance or the interfacial thermal resistance, $R_k$ is defined as
\begin{eqnarray}
R_k = \frac{\Delta T}{J_q} \label{eqn1}.
\end{eqnarray}
Also, the interfacial thermal conductance, $G_k$ can be defined as the inverse of $R_k$,
\begin{eqnarray}
G_k = \frac{1}{R_k} = \frac{J_q}{\Delta T} \label{eqn2}
\end{eqnarray}
where $J_q$ is the heat flux at the fluid-solid interface and $\Delta T = T_f - T_w$ is the difference between the temperature of the solid wall, $T_w$  and a fluid slab of some small thickness immediately adjacent to the solid wall, $T_f$.  For the graphene-water interface, the average thickness of the water is found to be about one molecular diameter \cite{alosious2020kapitza}.  
Here we are assuming that the time-dependent Kapitza kernel can be
expressed as an n-term Maxwellian memory function and is given by \cite{evans2008non}, 
\begin{eqnarray}
G_k\left(t\right)  = \sum\limits_{i = 1}^n {{k_i}} {e^{ - {\mu _i}t}}. \label{eqn3}
\end{eqnarray}
Here, $k_i$ and $\mu_i$ are coefficients in the Maxwellian memory
function and are related by, 
\begin{eqnarray}
{{\tilde C}_{T{J_q}}}\left( s \right) = \sum\limits_{i = 1}^n {\frac{{{k_i}}}{{s + {\mu _i}}}} {{\tilde C}_{{T{T}}}}\left( s \right). \label{eqn4}
\end{eqnarray}
${{\tilde C}_{T{J_q}}}$ and ${{\tilde C}_{TT}}$ are the Laplace transforms of the heat flux-temperature difference cross-correlation function and temperature difference autocorrelation function, respectively. Thus, for steady-state conditions ($s = 0$) we have,
\begin{eqnarray}
G_k  \equiv \tilde G_k \left( 0 \right) = \sum\limits_{i = 1}^n {\frac{{{k_i}}}{{{\mu _i}}}}. \label{eqn5}
\end{eqnarray}
Finally, the Kapitza resistance can be computed as,
\begin{eqnarray}
R_k = \frac{1}{G_k}.
\label{eqn6}
\end{eqnarray}
The Kapitza length can be defined as:
\begin{equation}
L_{k}=R_k\lambda
\label{eq:kl}
\end{equation}
where $\lambda$ is the thermal conductivity of either the solid or fluid phase. $L_{k}$ is similar to the slip length in fluid flow and can be defined as the additional thickness of material required to achieve the same heat transfer in place of the thermal resistance at the interface. This additional thickness can be measured either in the direction of solid or fluid from the interface as per convenience. The thermal conductivity of the solid phase is used in Eq.(\ref{eq:kl}) to determine the Kapitza length towards the fluid side and vice versa.
\par
The instantaneous heat flux, $J_q(t)$ and the temperature difference $\Delta T (t)$  can be extracted from the MD simulations with the help of the following equation  \cite{todd2017nonequilibrium,todd1995heat}
\begin{equation}
J_{q}(t)=J_{q}^{K}(t)+J_{q}^{\phi}(t),
\label{eqn7}
\end{equation}
where $J_{q}^{K}(t)$ is the kinetic term and $J_{q}^{\phi}(t)$ is the potential term of the heat flux. The reader is referred to the original paper for further details \cite{todd2017nonequilibrium,todd1995heat}.

\section{Methodology}
\begin{figure}[h!]
\centering
\includegraphics[width=1.0\textwidth]{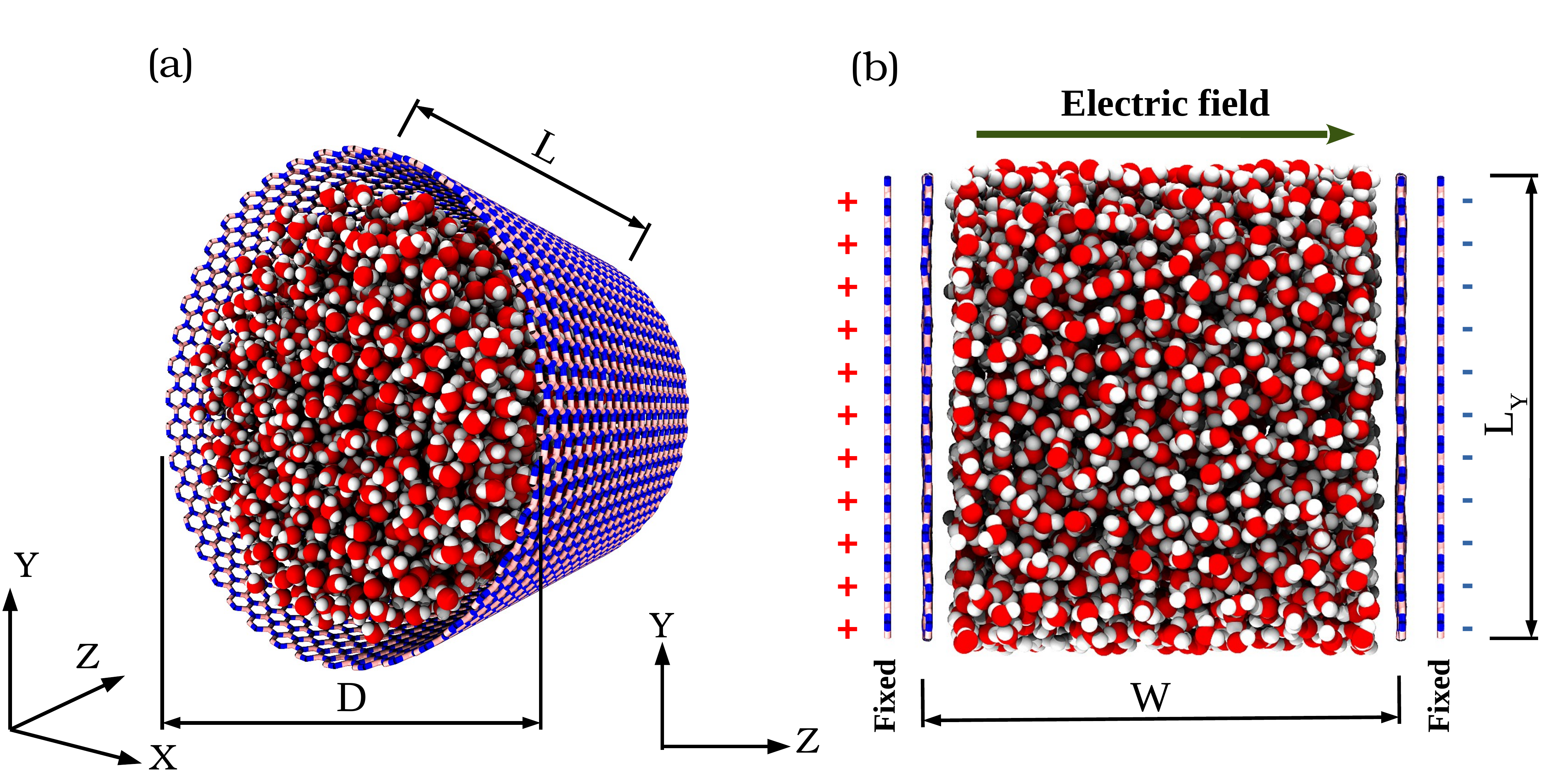}
\caption{Schematic depiction of the simulation systems. (a) Water confined in a hexagonal boron nitride nanotube. (b) Water confined in a planar hexagonal boron nitride nanochannel.}
\label{model}
\end{figure}

The simulation models for the planar and cylindrical hexagonal boron nitride (hBN)-water systems are shown in Figure \ref{model}. The planar system consists of a water block confined between two layers of parallel hBN sheets in which the outer layers were kept rigid to maintain a constant volume. The dimensions of the hBN sheets are $L_x$=$L_y$=4 nm, and the thickness of the water block, $W$=4 nm. Periodic boundary conditions were given in $x$ and $y$ directions, and confinement was in the $z$-direction. The hBN sheets were tethered to their initial position by fixing the centre of mass of the layers to their initial positions for ensuring a constant channel volume~\cite{alosious2020kapitza}.  The cylindrical system consists of water confined inside a hBN nanotube (hBNNT) of diameter, $D$ and length $L$=5 nm. Periodic boundary conditions were provided in all three directions. Different hBNNT diameters of 1.38 nm, 2.76 nm, 4.14 nm,  5.52 nm and 6.90 nm were chosen, which correspond to the chiralities of (10-10), (20-20), (30-30), (40-40) and (50-50) respectively. The diameter of an (m,n) hBNNT was calculated using the equation $d=\sqrt{3}l/\pi\sqrt{n^2+nm+m^2}$ where $l$=0.1446 nm is the  bond length of boron-nitride. Different parameters such as partial charge of hBN, salt concentration and external electric field were studied to understand the effect of electrostatic interactions on the Kapitza resistance. Boron nitrides (BN) are compounds having covalent–ionic bonds; hence binding polarity is a crucial factor influencing their physical characteristics. The dependence of observable parameters on effective static charges of component atoms is so intricate that it is nearly impossible to determine experimentally. The theoretically calculated atomic charges in boron nitrides are characterized by a large dispersion, leaving them practically unreliable. The absence of a clear distinction of the electron density between atoms of elements is the general explanation \cite{chkhartishvili3estimation}. In molecular mechanics force fields, partial atomic charges are utilized to calculate the electrostatic interaction energy. They are often utilized to get a qualitative understanding of a molecule's structure and reactivity. For BN, different partial charge values ranging from 0 to 1.05e were reported in various literatures \cite{won2008structure,chkhartishvili3estimation,won2007water,hilder2010validity}
The partial charge of the hBN was varied from 0.0 to $\pm$ 1.50$e$ to study the effect of partial charge. The effect of the electrolyte concentration was studied by varying the salt (NaCl) concentration from 0 to 1.0 mol/L. In addition to that, the impact of the external electric field was studied by varying the electric field strength from 0 to 0.1 V/\AA, applied in the direction perpendicular to the walls. The outer hBN layers on the left and right sides were provided with equally distributed positive and negative charges of the same magnitude, respectively. The outer hBN layers act as electrodes and develop an electric field in the $z$-direction (normal to the surface). The surface charge density was varied from 0 to 0.1375 $\mu$C/cm$^2$ to obtain an electric field in the range of 0 to 0.1 V/\AA, similar to previous MD simulations \cite{yenigun2019electric,luedtke2011dielectric,yen2012investigation}.
The range of electric fields selected here is higher than those typically used in experimental studies. However, transient pulse fields can be used as an alternative to achieve this range of electric field strengths experimentally.  The electro-freezing of water was experimentally investigated by different researchers using a pulse voltage applied through two electrodes immersed in water to create large electric fields \cite{braslavsky1998electrofreezing,petersen2006new}. The electric field between the two electrodes was reached in the range of 0.1 V/\AA, similar to the field strength used in this study.  In addition to that,  extreme field strengths up to 5V/\AA\ were also applied to study the structural changes of water due to electric fields by assuming that the dielectric breakdown of water will not occur.

The water molecules were modelled using the simple point charge  (SPC/E)~\cite{berendsen1987missing,wu2006flexible} water model due to its reliability, precision, and comparatively low computational cost. The long-range electrostatic forces were calculated using the particle-particle-particle-mesh (PPPM) \cite{hockney1988computer} solver with an accuracy of 1$\times$10$^{-5}$ and the water molecules were kept rigid using the SHAKE~\cite{ryckaert1977numerical}  algorithm. However, to use the PPPM method, periodic boundary conditions in all directions are required.  For the planar system, an Ewald summation technique with an extended volume ratio of 3.0 was used due to the confinement in the $z$-direction ~\cite{yeh1999ewald}. For the cylindrical system, void spacing of 5 nm in the $x$ and $y$ directions was provided to eliminate the errors due to electrostatic interaction between periodic images while implementing the PPPM solver \cite{ostler2017electropumping}.
An optimized Tersoff~\cite{lindsay2010optimized} potential was used to model the hBN nanosheet/tube interactions.  The pairwise interactions between all the atoms and ions were defined by adding the Lennard-Jones (L-J) and Coloumbic potentials and which is given by, 
\begin{equation}
U_{ij}=\sum_{i=1}^{N} \sum_{j=i+1}^{N} \Big(4 \varepsilon_{ij}\Big[\Big(\frac{\sigma_{ij}}{r_{ij}}\Big)^{12}  
- \Big(\frac{\sigma_{ij}}{r_{ij}}\Big)^{6}\Big] 
+ \frac{q_{i}q_{j}}{4\pi\varepsilon_{0}r_{ij}}\Big).
\label{lj}
\end{equation}
The values of the L-J interaction parameters are provided in Table~\ref{ljpara}. The L-J parameters for the cross-terms were calculated using the Lorentz-Berthelot mixing rules. 
\begin{table}[]
\centering
\begin{tabular}{@{}lllll@{}}
\toprule
Pair   & $\sigma$ (\AA) & $\epsilon$ (kcal/mol) & $q(e)$  & Reference                                                \\ \midrule
H-H    & 0.0            & 0.0                   & $+$0.4238 & Wu \textit{et al.} \cite{wu2006flexible}                 \\
O-O    & 3.166          & 0.1554                & $-$0.8476 & Wu \textit{et al.} \cite{wu2006flexible}                 \\
B      & 3.453          & 0.0949                & $+$1.05   & Won \textit{et al.} \cite{won2007water,won2008structure} \\
N      & 3.365          & 0.145                 & $-$1.05   & Won \textit{et al.} \cite{won2007water,won2008structure} \\
Na$^+$ & 2.160          & 0.3526                & $+$1.0      & Joung \textit{et al.} \cite{joung2008determination}      \\
Cl$^-$ & 4.831          & 0.0128                & $-$1.0      & Joung \textit{et al.} \cite{joung2008determination}      \\ \bottomrule
\end{tabular}
\caption{The L-J parameters used for modeling the hBN-water system.}
\label{ljpara}
\end{table}
The cut-off distance values for L-J potential and short-range Coulombic interactions are set as 1 nm. All the Molecular Dynamics (MD) simulations were performed by using the Large-scale Atomic/Molecular Massively Parallel Simulator (LAMMPS)~\cite{plimpton1995fast} package with a time step of 1.0 fs. Visual Molecular Dynamics (VMD)~\cite{HUMP96} was used to visualize the models.

Equilibrium molecular dynamics (EMD) simulations were performed to compute the Kapitza resistance at the hBN-water interface in planar and cylindrical geometry. Initially, the system was equilibrated under
a canonical (NVT) ensemble at a reference temperature of 300 K for a time period of 2.0 ns. Further, the system stability was checked by simulating under a microcanonical (NVE) ensemble for another 2.0 ns. Finally, the instantaneous heat flux, $J_q$(t) and temperature difference, $\Delta T(t)$, were extracted from another 5.0 ns production simulation by thermostatting the hBNNT only. The Kapitza resistance was calculated using the Equations \ref{eqn4}-\ref{eqn6} with a Maxwellian one term (n = 1) memory function \cite{alosious2019prediction}. The fluid slab thickness, $\Delta$ was taken as 3.165 \AA, which is the distance measured from the wall to the first density peak of water~\cite{alosious2020kapitza,alosious2021nanoconfinement}. All the calculated parameters were averaged over five independent simulations with distinct initial configurations.

\section{Results and discussions}
\begin{figure}[h!]
\centering
\includegraphics[width=1.0\textwidth]{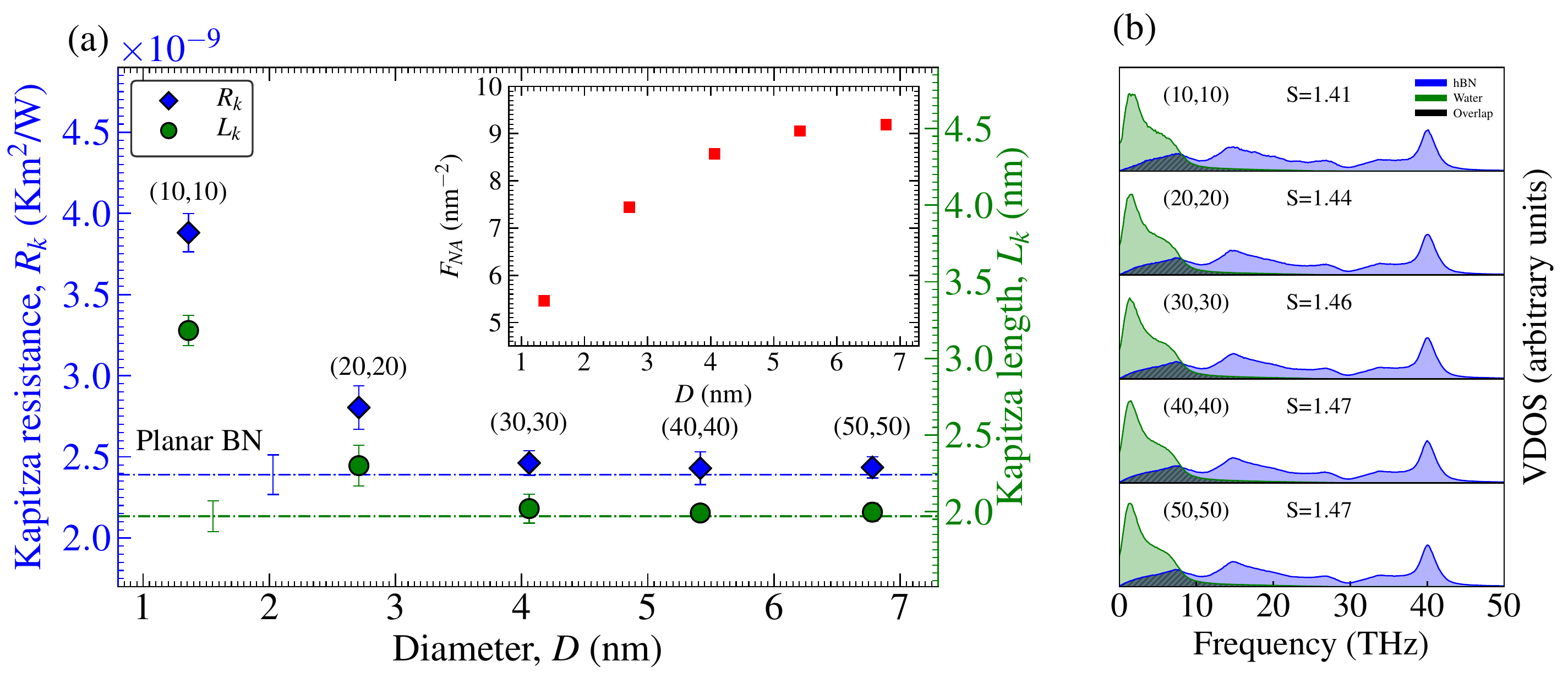}
\caption{Kapitza resistance/length as a function of the diameter of hBNNT. The horizontal dashed lines represents the Kapitza resistance/length
in a planar hBN-water system, and the inset shows the change in $F_{NA}$, with the diameter of the hBNNT. (b) VDOS (arbitrary units) and the overlap, S, of the hBNNT and the water slab for various hBNNT diameters. }
\label{dia}
\end{figure}
The computation of Kapitza resistance/length in a cylindrical nanoconfinement system using the NEMD simulation technique is challenging and complex and limited to a range of diameters. In our previous work, we have developed an EMD method to compute the Kapitza resistance in a CNT-water interface, and the results were validated with the NEMD method \cite{alosious2021nanoconfinement}. Results of both the EMD method and NEMD method are in excellent agreement. In this work, we also use the same EMD method to compute the Kapitza resistance in the hBN-water interface. Since the present study also uses the same geometry, this paper does not provide additional validation of the EMD method for the hBNNT-water interface.
Figure \ref{dia}a shows the effect of diameter of hBNNT on the Kapitza resistance/length. The dashed lines depict the Kapitza resistance/length of the planar hBN-water system (water confined in a hBN nanochannel). It is found that the Kapitza resistance reduces monotonically with an increase in the diameter of hBNNT and converges to a constant value similar to the CNT-water system. There is a 37.3\% reduction in the Kapitza resistance observed when the diameter varied from 1.38 nm to 6.9 nm, corresponding to the chiralities of (10,10) to (50,50), respectively. The inset shows the variation of area density factor, $F_{NA}$ as a function of diameter. $F_{NA}$ is defined as the ratio of the number of water molecules adjacent to the interface to the surface area of the hBNNT \cite{alosious2021nanoconfinement}. A slab of water adjacent to the wall is chosen for counting the number of molecules adjacent to the interface. The water slab thickness is taken as the distance from the wall to the first density peak of water since the first hydration shell determines the major part of the interfacial heat transfer. It is found that $F_{NA}$ increases with an increase in diameter and reaches a constant value for higher diameters. A higher value of $F_{NA}$ indicates that more water molecules will be present near the interface per unit surface area leading to a better heat transfer between hBNNT and water, thereby reducing the Kapitza resistance. This can be replaced by the number of molecules or the height of the first density peak for planar systems since the surface area remains the same. Even though the interactions between hBNNT and water remains the same, for smaller diameters, the slab volume to surface area is less (curvature effect), leading to a higher Kapitza resistance and this curvature effect vanishes after a particular diameter. Thus the increase in area density factor with an increase in diameter is the primary reason for reducing the Kapitza resistance. Furthermore, we have evaluated the mismatch of vibrational density of states (VDOS) between hBNNT and water to understand the interfacial heat transfer. The VDOS can be calculated using the Fourier transform of the velocity autocorrelation function given by, \cite{gao2019graphene,grest1981density}

\begin{equation}
P(f)=\frac{1}{\sqrt{2\pi}}\int_{-\infty}^{+\infty} {C\left( t \right)}e^{ift}dt,
\label{equationvdos}
\end{equation}
where $P(f)$ is the VDOS, $f$ is the phonon frequency, $C(t)$ is the velocity autocorrelation function given by $ C(t)= \langle v(t)v(0) \rangle / \langle v(0)v(0) \rangle  $ and $v(t)$ is the velocity of atom at time $t$.

Also, the VDOS overlap $S$, (generally considered in terms of an arbitrary unit) of the  water and the hBNNT was calculated using the equation \cite{chen2009tunable},
\begin{eqnarray}
S= \frac{\int_0^{\infty}P(f)_{hBNNT}P(f)_{water}\,df}{\int_0^{\infty}P(f)_{hBNNT}\,df.\int_0^{\infty}P(f)_{water}\, df}
\label{vdos}
\end{eqnarray}
The energy carriers in water are molecules, and the phonon dynamics in bulk liquids are not well defined. However, under extreme confinements, surface charges, electric fields etc., a high ordering of water (solid-like structure) near the interface is possible. Therefore, here the VDOS overlap is calculated to show that it doesn't influence the thermal transport in a solid-fluid interface, unlike a solid-solid interface.
Figure \ref{dia}b shows the VDOS and its overlap, $S$ for hBNNT and water. There is no noticeable change in $S$ with an increase in the diameter, indicating the vibrational coupling between hBNNT and water are unaltered with a change in diameter. Thus the reduction in the Kapitza resistance with an increase in diameter can be attributed to the increased area density factor, similar to the CNT-water system \cite{alosious2021nanoconfinement}.
\begin{figure}[h!]
\centering
\includegraphics[width=0.75\textwidth]{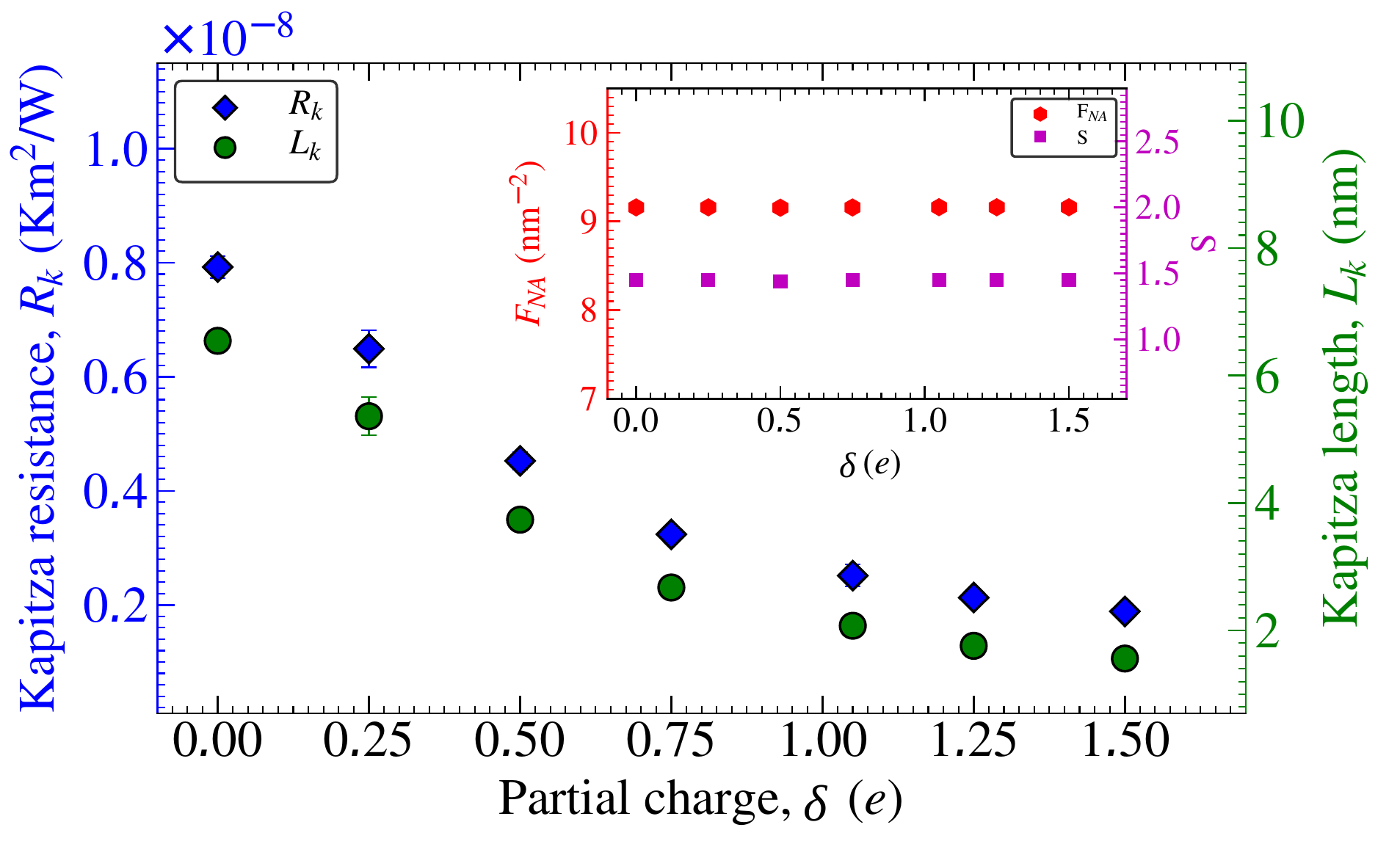}
\caption{Kapitza resistance/length as a function of the partial charge of a (40,40) hBNNT. The inset shows the effect of area density factor $F_{NA}$ and VDOS overlap, $S$ }
\label{charge}
\end{figure}

We further examine how the partial charge of hBNNT can influence the interfacial heat transfer in the present system.  Figure \ref{charge} shows the effect of the partial charge of a (40,40) hBNNT on the Kapitza resistance/length, and the inset shows the corresponding values of the area density factor $F_{NA}$ and the VDOS overlap, $S$. The partial charge on the hBNNT was varied between 0 to $\pm$ 1.50$e$ to obtain a range of values higher and lower than the actual value. It is found that the Kapitza resistance decreases with an increase in partial charge as expected. The Kapitza resistance was reduced to nearly 71.2\% by increasing the partial charge from 0 to $\pm$ 1.50$e$.  A meagre value of Kapitza resistance for the hBNNT-water interface compared to the CNT-water interface is primarily due to the presence of partial charges on the hBNNT. To examine how the Kapitza resistance is affected by partial charge, we have calculated the area density factor, $F_{NA}$, and the VDOS overlap $S$ for different partial charges. 
From the inset of Figure \ref{charge}, we can see that the area density factor is almost independent of the partial charges. Alexeev \textit{et al.} \cite{alexeev2015kapitza} reported that the Kapitza resistance is primarily dependent on the water density profile near the interface and is inversely proportional to the height of the first density peak in a water-graphene system. However, a constant value of the area density factor indicates that the density profile near the interface is not changing with the addition of partial charge (since the diameter of hBNNT is the same, $F_{NA}$ and density peaks are proportional). Similar observations were previously reported in graphene and hBN nanoconfinement systems such that the partial charge or surface charge (zero net charge) doesn't influence the density profile of water \cite{ma2018ordered,tocci2014friction,won2008structure}. This indicates that the reduction in Kapitza with partial charges are not related to the change in the structure of the water near the interface. In addition to that, the VDOS overlap between hBNNT and water, $S$ is also nearly independent of the partial charges indicating that the reduction in Kapitza resistance does not originate from the change in vibrational coupling between water and hBNNT. 

Furthermore, hydrogen-bond dynamics were analyzed to understand the physical mechanism behind the dependency of Kapitza resistance on partial charge. A hydrogen bond (H-bond) is essentially an electrostatic force of attraction between a hydrogen (H) atom, which is covalently bonded with a more electronegative atom or group (donor D), and another electronegative atom having a lone pair of electrons (acceptor A) \cite{arunan2011definition}. 
The hydrogen bonds in molecular models are determined by a geometric criterion that includes the distance between the hydrogen and acceptor atoms (H$\cdots$A) and the donor-hydrogen-acceptor ($\chemfig[atom sep=2.2em]{D-H}\cdots$A) planar angle value. In the hBN-water system, two types of H-bonds are possible; one is the bond between a partially positive hydrogen atom in a water molecule and a partially negative oxygen atom in a different water molecule ($\chemfig[atom sep=2.2em]{O-H}\cdots$O) and another is between a partially positive hydrogen atom in a water molecule and partially negative nitride in the hBN ($\chemfig[atom sep=2.2em]{O-H}\cdots$N). Here, two molecules were considered H-bonded if (H$\cdots$A) < 3.5 \AA\ and ($\chemfig[atom sep=2.2em]{D-H}\cdots$A) > 140$^\circ$ \cite{cicero2008water,prasad2018water}. The H-bond lifetime, $\tau$ can be used to analyze the dynamic behaviour of H-bonds using the time autocorrelation function given by \cite{rapaport1983hydrogen}, 

\begin{equation}
C_c(t)=\left\langle \frac{\sum h_{ij}(t_0)h_{ij}(t_0+t)}{\sum h_{ij}(t_0)^2}\right\rangle
\label{eqntime}
\end{equation}

Where $h_{ij}$ is a binary measure of whether pair $ij$ forms a H-bond or not such that, $h_{ij}$ = 1, or  $h_{ij}$ = 0. Here we adopted the continuous lifetime definition, in which once the H-bond is broken, it is always considered to be broken even if the bond reforms afterwards. The H-bond lifetime $\tau$ can be defined as \cite{rapaport1983hydrogen}, 
\begin{equation}
\tau=\int_0^{\infty} C_c(t).
\label{eqnlife}
\end{equation}

\begin{figure}[h!]
\centering
\includegraphics[width=1.0\textwidth]{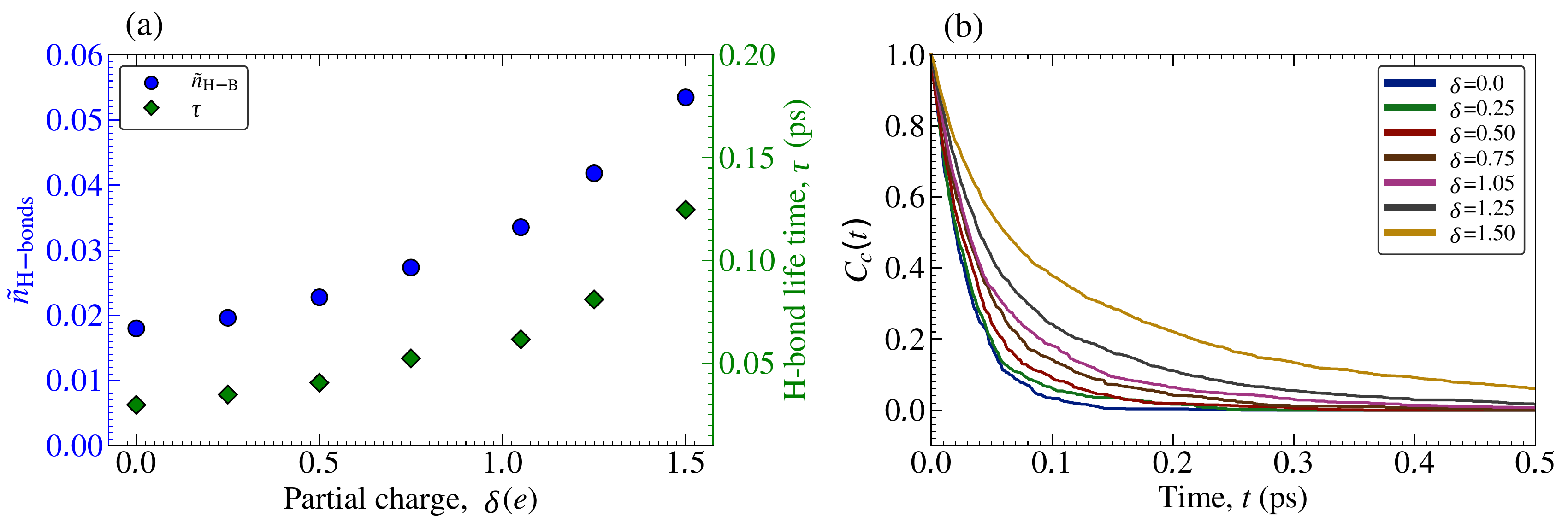}
\caption{(a) Number of water-hBNNT hydrogen bonds per molecule and water-hBNNT hydrogen bond life time as a function of partial charge. (b) Water-hBNNT hydrogen-bond time autocorrelation function for different partial charges. }
\label{hbond}
\end{figure}

We have calculated the number of hydrogen bonds per molecule, $\tilde{n}_{\mathrm{H-B}}$ and hydrogen bond lifetime, $\tau$ for water-water ($\chemfig[atom sep=2.2em]{O-H}\cdots$O) and water-hBN ($\chemfig[atom sep=2.2em]{O-H}\cdots$N) hydrogen bonds. The number of hydrogen bonds per water molecule is around 3.18, and the H-bond lifetime is 0.30 ps for the water-water hydrogen bonds.  Both the values remain the same even after increasing the partial charge of the hBNNT. This indicates that the water-water hydrogen bond dynamics are not affected by the partial charge of the hBNNT. In contrast, the H-bond between hBN and water is found to be strongly dependent on the partial charge. Figure \ref{hbond}a shows the number of hydrogen bonds per molecule and H-bond lifetime for water-hBN hydrogen bonds. The hydrogen bond between water and hBN is only possible in a single layer of water near the interface; therefore, the number of hydrogen bonds per molecule will be meagre compared to the water-water H-bonds. It is found that the number of hydrogen bonds per molecule increases with an increase in the partial charge of the hBNNT. This indicates that the partial charges are favourable for forming more H-bonds between water and the hBNNT, leading to a better heat transfer at the interface. Figure \ref{hbond}b shows the H-bond time autocorrelation function for different partial charges. The correlation function converges quickly for lower partial charges, and as we increase the partial charge, the convergence time increases, leading to a higher H-bond lifetime. Thus it is evident that the H-bond lifetime increases with an increase in partial charges leading to a stronger H-bond at the interface. From the H-bond analysis, we can infer that the partial charge of hBNNT strongly influences the hydrogen bonding dynamics at the interface leading to higher Coloumbic interaction between water and hBNNT. Thus the reduction in the Kapitza resistance with an increase in partial charge of the hBNNT can be attributed to the increased hydrogen bonding between water and hBNNT. 

The partial charge does not influence the structuring and ordering of water near the interface, and the density of water is unaltered with the addition of partial charge in boron nitride. Similar observations were previously reported in graphene and BN nanoconfinement systems such that the partial charge or surface charge (zero net charges) doesn't influence the density profile of water \cite{ma2018ordered,tocci2014friction,won2008structure}.
However the density profile changes when a net charge density is provided on the solid. This shows that different charge arrangements lead to different mechanisms at the interface. Even though the magnitude of the heat transfer rate across the interface would increase due to the heat flux contribution from the electrostatic force, the hydrogen analysis is carried out to understand the physical mechanism underlying interfacial thermal transport.
\begin{figure}[h!]
\centering
\includegraphics[width=0.75\textwidth]{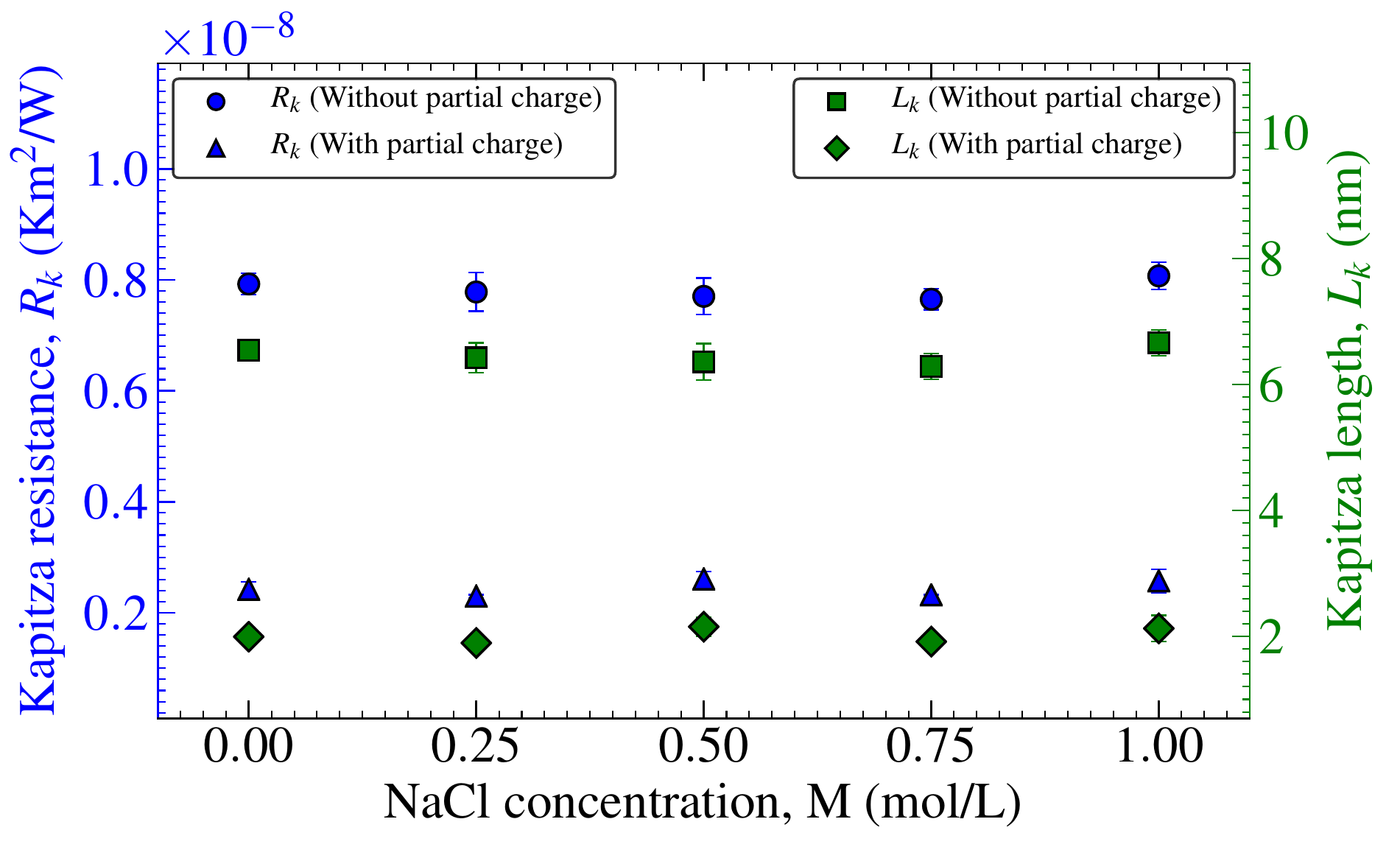}
\caption{Kapitza resistance/length as a function of NaCl concentration for hBNNT with/without partial charge ($\delta =1.05 e$). }
\label{electrolyte}
\end{figure}

Next, we examined whether the addition of salt ions can influence the interfacial thermal transfer characteristics. Figure \ref{electrolyte} shows the effect of the NaCl concentration on the Kapitza resistance/length for a (40,40) hBNNT with/without partial charges. The concentration of the NaCl electrolyte varied from 0 to 1.0 mol/L. For the hBNNT without partial charges, we can see that the Kapitza resistance is independent of the salt concentrations. Similarly, for hBNNT with partial charges, the Kapitza resistance reduces due to the partial charge but remains unchanged with increased concentration. 
This indicates that the addition of NaCl does not influence interfacial thermal transport in both systems. To further understand the underlying mechanism, we examined the density profiles of the hBNNT-electrolyte system.
\begin{figure}[h!]
\centering
\includegraphics[width=0.85\textwidth]{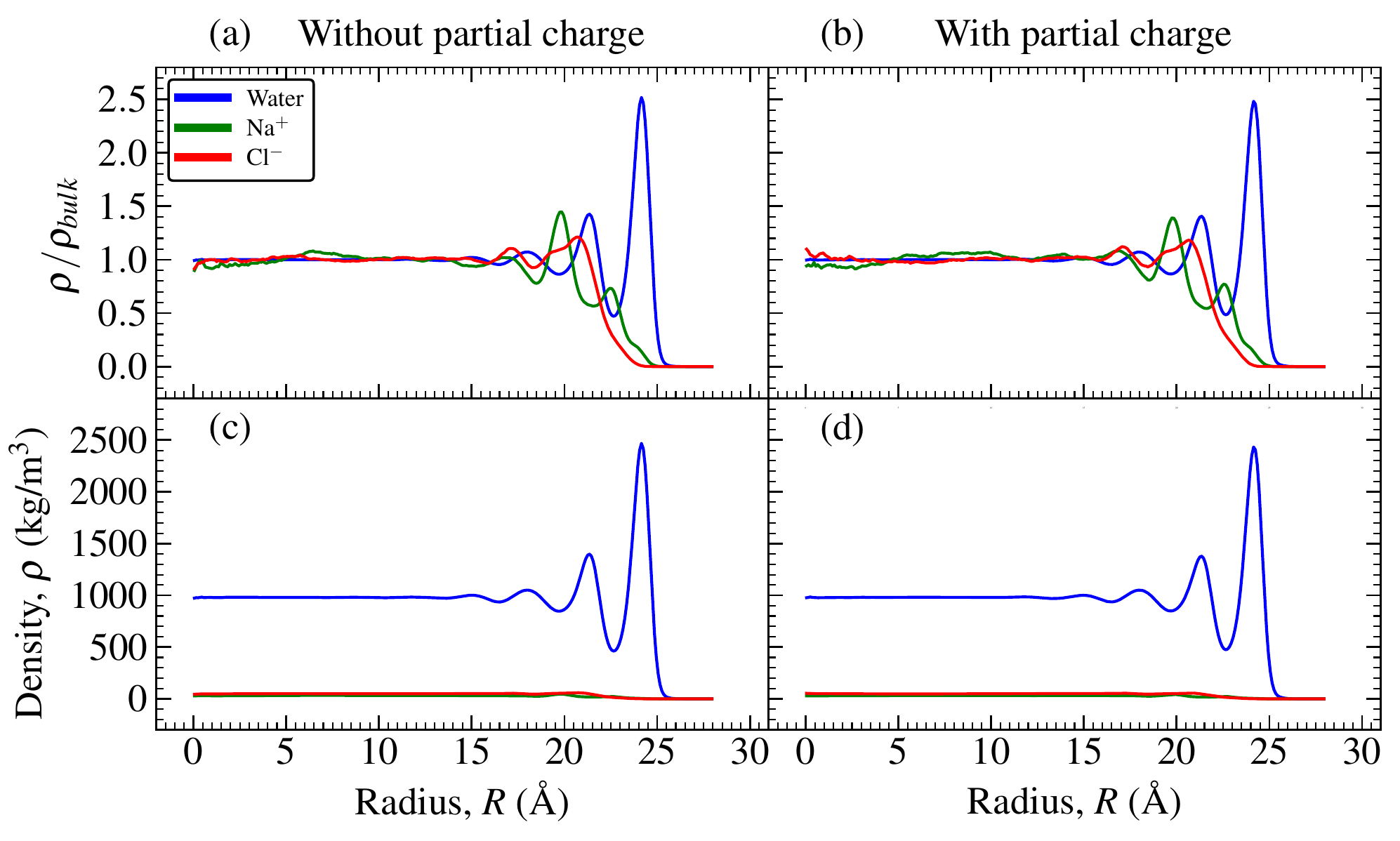}
\caption{Density profile of water and ions for 1.0 mol/L NaCl concentration.  (a),(b) Normalized density profile of water and ions with/without partial charge. (c),(d) Mass density profile of water and ions with/without partial charge.}
\label{density}
\end{figure}
 Figure \ref{density} shows the density profiles for different conditions with 1.0 mol/L NaCl concentration. The normalized density profile for water and ions for the highest concentration without and with partial charges of the hBNNT is shown in Figure \ref{density}a and \ref{density}b, respectively. Similarly, Figure \ref{density}c and \ref{density}d show the absolute density profile for both systems. The Kapitza resistance is a strong function of the height of the first density peak of water \cite{alexeev2015kapitza}. We observed that the height of the density peak is unaltered with the addition of NaCl even at the highest concentration. From Figure \ref{density}a, we can see the normalized density fluctuation of Na$^+$ and Cl$^-$ ions along with the water molecules for the hBNNT without partial charges. The structure of water governs the density profile of the ions at the interface in such a way that the ions are concentrated in the valleys of the water density profile near the interface and equally distributed in the bulk portion. Since the Kapitza resistance is mostly dependent on the first layer, the concentration of ions after the first density peak do not contribute much to the interfacial thermal transport. Thus the addition of NaCl even at higher concentrations does not influence the Kapitza resistance. Furthermore, Figure \ref{density}c shows the absolute density profile for the same system. We can see that, compared to the density of water, the density of both Na$^+$ and Cl$^-$ ions are very negligible. Thus the contribution from NaCl for interfacial heat transfer will be meagre compared to the contributions from water molecules. So combining both, we can conclude that the concentration-independent Kapitza resistance is due to the relatively minor contribution from NaCl ions than water molecules and the position of ions far from the interfacial region. In addition to that, we have examined the effect of salt concentration when the hBNNT is applied with partial charges. When the partial charges are provided, the Coulombic interaction between water and hBNNT will increase, thereby reducing the Kapitza resistance. Even though the partial charge of hBNNT and the charge of ions also enhance the Coulombic interaction, it is significantly less compared to the water and not enough to make an additional influence on interfacial thermal transport.  Thus the Kapitza resistance is still found to be unaltered with the addition of NaCl. Figure \ref{density}b and \ref{density}d show the normalized and absolute density profile of the electrolyte. Even after applying the partial charges to the hBNNT, the density profiles nearly match the hBNNT without partial charges. Thus, adding NaCl to water alone does not affect the interfacial thermal transport and Kapitza resistance in both systems (with/without partial charge).

\begin{figure}[h!]
\centering
\includegraphics[width=0.75\textwidth]{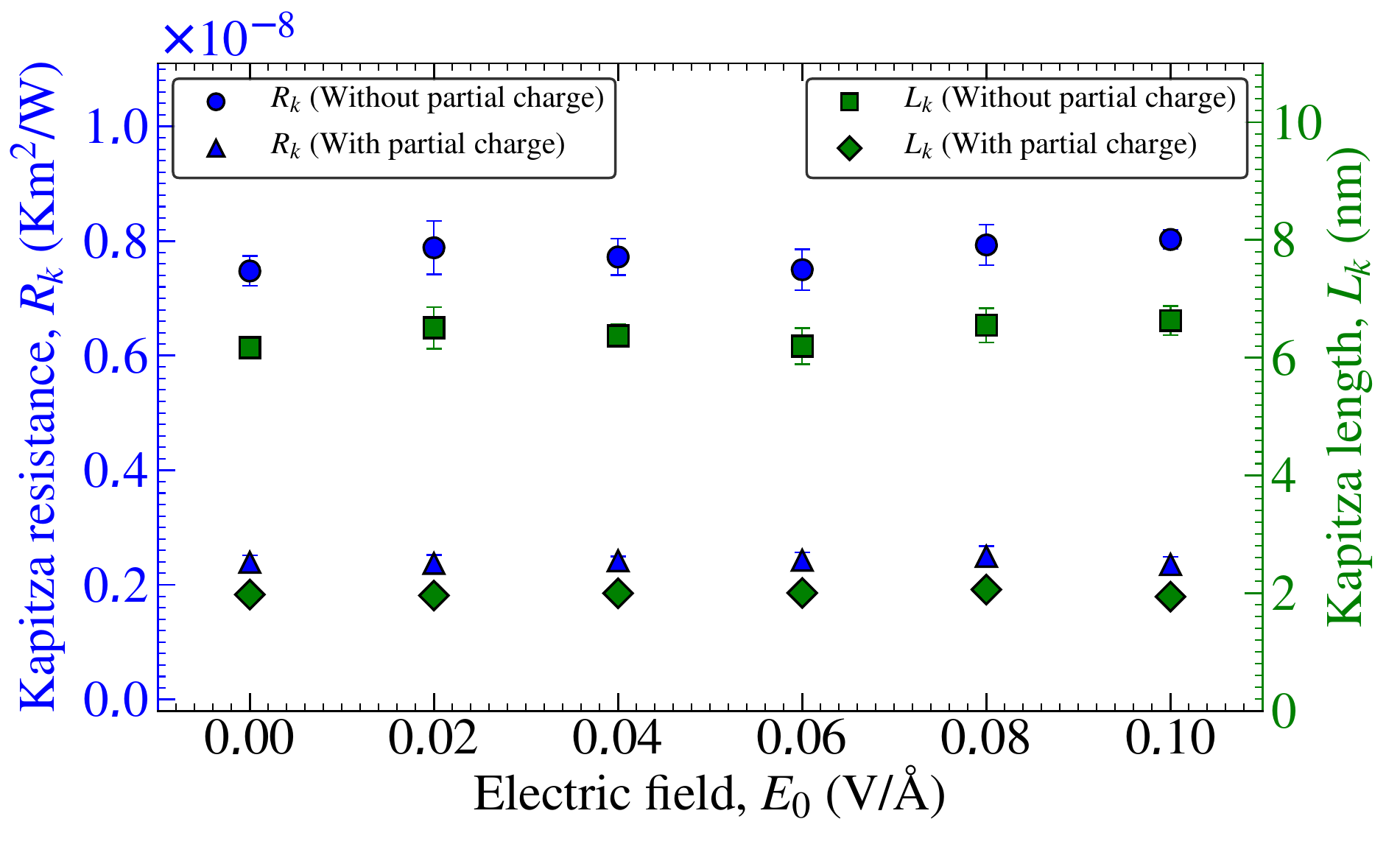}
\caption{Kapitza resistance/length as a function of low electric field in a planar hBN with/without partial charge ($\delta =1.05 e$).}
\label{field}
\end{figure}

Applying an electric field to water can influence its properties due to the polar nature of water. Here we examine the effect of electric field on interfacial thermal transport in a planar hBN-water nanochannel. An electric field perpendicular to the interface (along the $z$ direction) is applied to the system by providing positive and negative surface charges on the outer layers of hBN as shown in Figure \ref{model}.
The positive uniform charge density on the left side acts as an anode, and the negative uniform charge density with the same magnitude acts as a cathode creating a uniform electric field from left to right. The electric field varied from 0 to 0.1 V/\AA\ by changing the surface charge density from 0 to 0.1375 $\mu$C/cm$^2$. 

The polarization of water in response to the applied electric field, $E_0$, in the $z$ direction creates an induced electric field given by,
\cite{glosli1996molecular}
\begin{equation}
E_p(z)=\frac{\int_{-\infty}^z\rho_q(z')dz'}{\epsilon_0},
\label{field1}
\end{equation}
where $\rho(z')$ is the charge density distribution along the $z$ direction and $\epsilon_0$ is the dielectric permittivity of vacuum. The spatial distribution of charge density $\rho(z)$ is defined as
\begin{equation}
\rho(z)=\frac{1}{L_xL_y}\left\langle \sum_i^N q_i\delta(z-z_i)\right\rangle, 
\label{field2}
\end{equation}
where $q_i$ is the charge of the $i$th atom. The net electric field is given by,\cite{yeh1999dielectric}
\begin{equation}
E(z)=E_0+E_p(z).
\label{field3}
\end{equation}
At the interface, the induced polarization field, $E_p$ is excessively higher than the applied electric field $E_0$  \cite{varghese2019effect}. This causes the overscreening of the applied field by the polarization field. For an applied field strength of $E_0$=0.1 V/\AA, the net electric field in the bulk portion of the channel is obtained as $E$=0.0016 V/\AA.

Figure \ref{field} shows the effect of electric field on the Kapitza resistance/length in the planar hBN-water system with and without partial charges. The Kapitza resistance is found to be nearly independent of the electric field in the range of 0 to 0.1 V/\AA. Even though the applied fields are  higher compared to experimental values, it is not enough to make any structural changes in the water. As stated earlier, the density profile and the height of the first peak is the primary deciding factor controlling the interfacial heat transfer. Here we found that the density profile coincides with each other for all the applied external electric fields (not shown here). The electric field can influence the orientation of water molecules \cite{ren2015interfacial,varghese2019effect}; however, it does not make any impact on the interfacial heat transfer and the Kapitza resistance at this range of field strengths.

Substantial electric field strength is required to induce any structural changes in water. Numerous studies using MD simulations have been reported to understand the structural behaviour of water under extreme magnitudes of electric fields as high as 5V/\AA \cite{xia1995electric,yeh1998structure,hu2011response,zhang2014influence,nie2015role}.  Applying such high electric fields in experimental setups are nearly impossible due to several constraints such as difficulty in achieving higher fields and the dielectric breakdown of water. It has been reported that the dielectric breakdown of water is in the order of 10$^{-3}$ V/\AA\ and 10$^{-2}$ V/\AA\ for millimetre and micrometre scale volumes of water, respectively \cite{stygar2006water,song2010high}. This indicates that nanoconfinement of water could enhance the dielectric strength of water further; however, it is yet to be experimentally validated. 
Also, the MD simulations have not reported the dielectric breakdown of  water even in this range of applied electric fields.  
However, we cannot completely ignore the possibility that the observations obtained from MD simulations under these extreme electric field strengths are simulation artefacts. Nevertheless, to study the structural changes of water due to electric fields, we applied extensive electric field strengths up to 5 V/\AA\ by assuming that the dielectric breakdown of water will not occur. When the applied electric field exceeds the dielectric strength of water, the water will suddenly become an electrical conductor, and electric current flows through it. Predicting the thermal properties and interfacial thermal transport during the dielectric breakdown is challenging and is not attempted in this study. 
\begin{figure}[h!]
\centering
\includegraphics[width=0.9\textwidth]{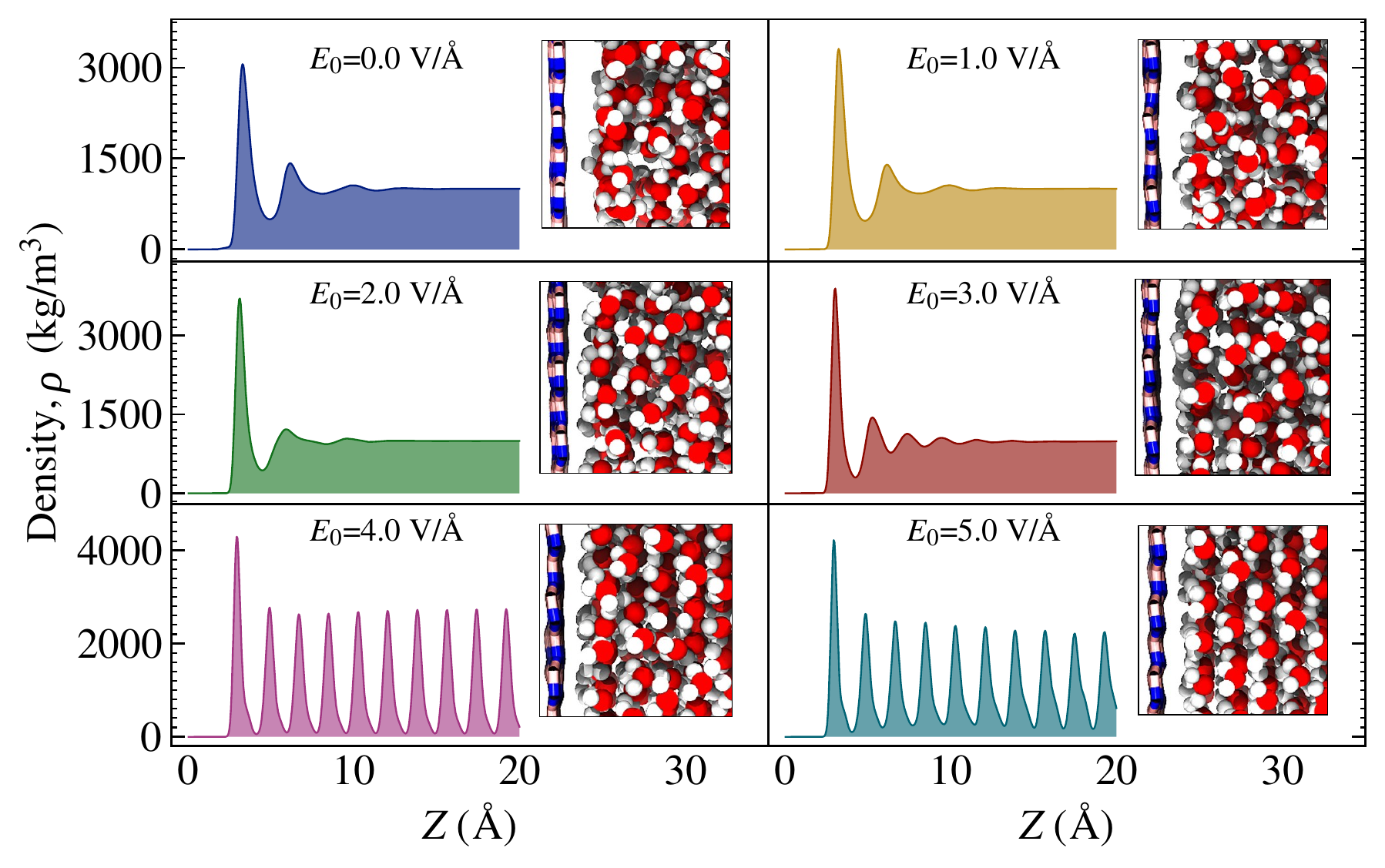}
\caption{Evolution of water density profile and snapshots of water molecules for different electric fields.}
\label{density_field}
\end{figure}

Figure \ref{density_field} shows the evolution of the water density profile and snapshots of water molecules with increasing electric field strength. While increasing the electric field, the water molecules start to form an ordered crystalline structure is known as electro-freezing \cite{zangi2004electrofreezing,yan2012molecular,zhu2014phase}. The density profile shows that the electro-freezing phenomenon occurs at about 4 V/\AA\ applied field strength. For zero and lower electric field strengths, we can see from the snapshots that the water molecules are randomly oriented, and when the field strength is increased, the ordering starts to occur. A highly ordered crystalline structure of water molecules starts to be visible from a field strength of 4 V/\AA\ and above. 
A similar range of electric field strengths at which electro-freezing occurs has also been reported in previous MD simulations of confined water
\cite{yenigun2019electric,celebi2017electric}.
In contrast to that, Yenigun and Barisik \cite{yenigun2019electric} reported electro-freezing of confined water at a lower field strength of 0.05 V/\AA. However, the electric field they have mentioned is the effective field inside the water by considering the relative permittivity of water and not the external applied electric field. From the surface charge density they have used, it is evident that the effective external field is about 4 V/\AA\ which is nearly 80 times higher than the effective field (since the relative permittivity of water, $\epsilon_r$ $\approx$ 80). The electro-freezing phenomenon in subcooled water has been widely reported in experimental studies \cite{sivanesan1991ice,yang2015ice,carpenter2015electrofreezing,zhang2016role}. However, the room temperature electro-freezing has not been experimentally detected due to different constraints in achieving extreme electric fields. Experiments are yet to validate the high ordering and crystallization of water molecules under intense electric fields at room temperature observed in MD simulations. Therefore the following results and analysis are strictly based on the assumptions that the dielectric breakdown of water does not happen and that room temperature electro-freezing is possible.  

\begin{figure}[h!]
\centering
\includegraphics[width=0.75\textwidth]{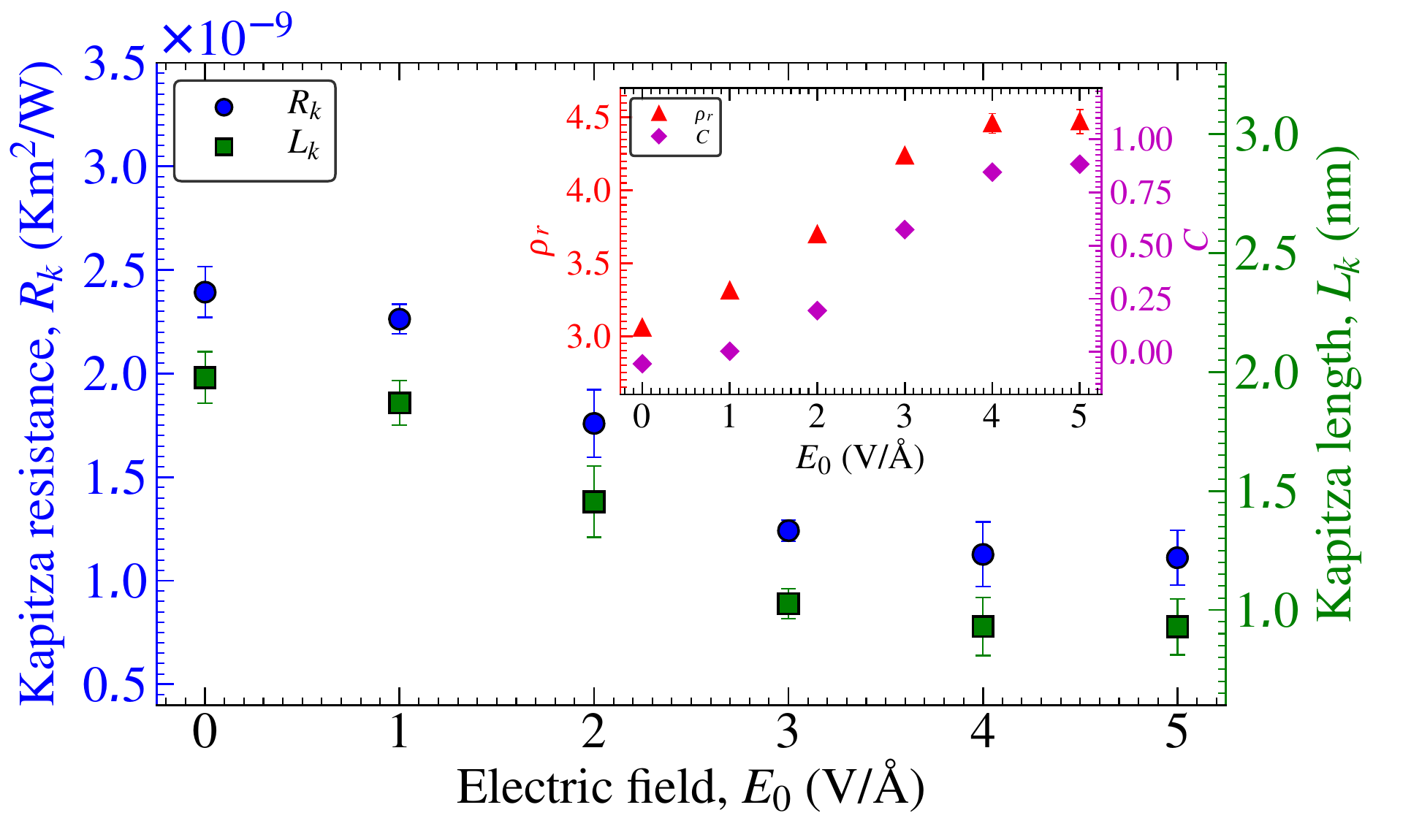}
\caption{Kapitza resistance/length as a function of high electric fields in a planar hBN-water system. The inset shows the reduced water density peak and order parameter as a function of electric field.}
\label{order}
\end{figure}

Figure \ref{order} shows the effect of electric field on the Kapitza resistance/length in the planar hBN-water system with partial charges, and the inset shows the change in order parameter and reduced water density peak for different electric field strengths. We found that the Kapitza resistance reduced by about 53.5\% when the electric field increased until 4 V/\AA. With a further increase in the field strength, the Kapitza resistance remains almost constant.  
Thus with the increase in the electric field, the ordering of water molecules commences, which reduces the Kapitza resistance, and once electro-freezing occurs, the Kapitza resistance remains constant with further increase in the electric field. To quantify the ordering of water molecules with the electric field, we computed the order parameter \cite{de1993physics}, which is defined as,
\begin{equation}
C=\frac{1}{2}\langle (3cos^2\theta-1)\rangle
\label{eqnorder}
\end{equation}
where C = 0 for random alignment, C = 1 for parallel alignment and  C = -1/2 for perpendicular alignment of molecular dipoles. The angle $\theta$ is defined by cos $\theta$  = $\hat{\textbf{\textit{n}}}.\hat{\textbf{\textit{E}}}$, where $\hat{\textbf{\textit{E}}}$ is the direction of applied field and $\hat{\textbf{\textit{n}}}$ is the direction of dipole moment.
From the inset of Figure \ref{order} we can observe that when the electric field increases from 0 to 4 V/\AA, the order parameter increases monotonically and converges to a steady value once the electro-freezing started to appear. Similarly, the reduced first water density peak ($\rho_r$=$\rho_{max}/\rho_{bulk}$) also increases with an increase in the electric field.  When the electric field is applied, due to the positive surface charge on the left hBN layer, the oxygen atoms with negative partial charge will orient towards the left wall, and the hydrogen atoms with positive partial charge will orient towards the negatively charged left hBN layer.  This orientation will increase with an increase in electric fields and be reflected in the value of the order parameter. Moreover, the ordering of water molecules towards the charged surface intensifies the layering effect, which increases the height of the first water density peak with an increase in the electric field. Combining the above observations, we can conclude that the reduction in Kapitza resistance with electric field strength is mainly due to the higher first water density peak and the ordering of water molecules leading to more aggregation of water molecules at the interface.

\section{Conclusions}
We have investigated the effect of different types of electrostatic interactions on the Kapitza resistance in cylindrical and planar hexagonal boron nitride-water systems using MD simulations. The cylindrical system consists of water confined in a hexagonal boron nitride nanotube (hBNNT), and the planar system consists of water confined between parallel hexagonal boron nitride (hBN) nanochannel. The calculation of Kapitza resistance was carried out using our previously developed EMD method due to the limitations of the NEMD method in cylindrical geometries. Our EMD method utilises the correlation functions of instantaneous heat flux and temperature difference data at the interface and provides correlation time-independent results.  We examined the effect of different factors affecting the Kapitza resistance, such as the diameter of hBNNT, partial charge, electrolyte concentration, and electric field strength.  
The Kapitza resistance monotonically decreases with an increase in diameter of the hBNNT and nearly converges to the value of the planar system, precisely similar to the trend in a CNT-water system. The Kapitza resistance was reduced to almost 37.37\% when the diameter increased from 1.38 nm to 6.9 nm primarily due to the increase in area density factor, $F_{NA}$ and the overlap of VDOS between water and hBNNT. The primary reason for a lower value of Kapitza resistance for hBNNT-water than the CNT-water interface is due to the presence of partial charges on hBNNT. A reduction in 71.2\% in Kapitza resistance was found when the partial charge increased from 0 to $\pm$ 1.50$e$. The partial charge does not influence the structuring and ordering of water near the interface. Whereas the hydrogen bonding between water and hBN wall increases with higher partial charges leading to a better heat transfer, thereby reducing the Kapitza resistance. Furthermore, we found that the addition of NaCl to water doesn't impact interfacial heat transfer. The Kapitza resistance is nearly independent of the electrolyte concentration since the contribution of electrostatic interaction of both Na$^+$ and Cl$^-$ ions are significantly less compared to the water molecules. Finally, the effect of the electric field on Kapitza resistance was studied by providing a uniform surface charge to the left and right hBN layers. The Kapitza resistance is found to be nearly independent of the electric field within the experimentally achieved range of 0 to 0.1 V/\AA. To study the structural changes of water due to electric fields, extensive electric fields up to 5 V/\AA\ were applied by assuming that the dielectric breakdown of water will not occur. The Kapitza resistance reduced by about 53.5\% when the electric field increased until electro-freezing occurs at approximately 4 V/\AA.  When the electric field is applied, the ordering of water molecules towards the charged surface intensifies the layering effect, which leads to an increase in the height of the first water density peak, thereby increasing the interfacial thermal transport. These findings provide an insight into the contributions of electrostatic interactions on Kapitza resistance, which helps to manage the interfacial thermal transport in various nanoscale systems.

\section*{Acknowledgements}

The authors are grateful for the computational resources provided by the P. G. Senapathy Center for Computing Resources at IIT Madras (AQUA supercluster) and the Swinburne supercomputing OzSTAR facility.
Authors S.P.S and S.A gratefully acknowledge the financial support received from the Ministry of Education, Government of India, 
under MHRD-STARS (Project Number: STARS1/336) to carry out a part of this work.  This research is also supported by the Indian Institute of Technology Madras to the Micro Nano-Bio Fluidics Group under the funding for the Institutions of Eminence scheme of Ministry of Education, Government of India [Sanction. No: 11/9/2019-U.3(A)].
\bibliography{reference}

\end{document}